\documentclass[letterpaper, 10 pt, conference]{ieeeconf}  

\IEEEoverridecommandlockouts                              

\usepackage{graphics} 
\usepackage{epsfig} 
\usepackage{mathptmx} 
\usepackage{times} 
\usepackage{amsmath} 
\usepackage{amssymb}  
\usepackage{algorithm}
\usepackage[noend]{algpseudocode}
\usepackage{textcomp}
\newtheorem{proposition}{Proposition}
\newtheorem{assumption}{Assumption}

\DeclareMathOperator*{\argmax}{arg\,max}

\usepackage{soul}
\usepackage{color}

\title{\LARGE \bf
Distributed Bio-inspired Humanoid Posture Control
}

\author{Vittorio Lippi$^{1}$, Fabio Molinari$^{1}$, Thomas Seel$^{1}$
\thanks{$^{1}$Technische Universit{\"a}t Berlin, Fachgebiet Regelungssysteme, Sekretariat EN11, Einsteinufer 1, D-10587 Berlin, Germany {\tt\small vittorio.lippi@tu-berlin.de, molinari@tu-berlin.de, thomas.seel@tu-berlin.de} .}			
				}

\begin{document}

\maketitle
\thispagestyle{empty}
\pagestyle{empty}

\begin{abstract}

This paper presents an innovative distributed  bio-inspired posture control strategy for a humanoid,
	employing a balance control system DEC (Disturbance Estimation and Compensation).
	Its inherently modular structure  
	could potentially lead to conflicts among modules, 
	as already shown in literature.
	A distributed control strategy is presented here,
	whose underlying idea is to let only one module at a time
	perform balancing, whilst the other joints are controlled to
		be at a fixed position.
	Modules agree, in a distributed fashion,
	on which module to enable, by iterating
	a max-consensus protocol.
	Simulations performed with a triple inverted pendulum model show that 
	this approach	limits the conflicts among modules
	while achieving the desired posture and allows for saving energy while performing the task. This comes at the cost of a higher rise time.
  
\end{abstract}

\section{Introduction}
\subsection{Overview}
Humanoid legged robots require bipedal balancing as base requisite to perform tasks. 
Since, at the state of the art, humans are still considered superior to robots at controlling posture \cite{NoriPetersPadoisEtAl2014},
there has been much interest
in implementing human inspired humanoid 
control systems. 
Nowadays, human-likeness of 
bipedal control is an ongoing research topic, see, e.g., \cite{torricelli2014benchmarking,lippibenchmarking,10.3389/fnbot.2018.00021}. 
The model proposed in this paper is the so-called
DEC (disturbance estimation and compensation) \cite{Mergner2010}, 
which is based on a neurological model of human posture. 
This model uses sensor fusion-derived 
internal reconstructions of the external 
disturbances affecting body posture.
Issues of a modular control system
are conflicts among joints, as extensively shown in \cite{ott2016good}.
In this work, 
we show how distributed 
control theory can be applied 
to such a framework.
In fact, a discrete-time max-consensus 
algorithm is used to define priorities 
between DEC modules in a distributed way,
so that conflicts can be prevented. 
Furthermore, this solution is compatible with  
plug-and-play frameworks, 
which may be relevant for future system reconfiguration.
\subsection{Notation}
In the remainder of this paper,
$\mathbb{N}$ and $\mathbb{R}$
denote, respectively,
the set of positive integers and the set of real numbers.
Nonnegative integers and nonnegative real numbers are
$\mathbb{N}_{\geq0}$ and $\mathbb{R}_{\geq0}$.
The indicator function is defined as follows
\begin{equation*}
    \mathbb{I}(x)=
    \begin{cases}
        1 \quad\text{if x is true}\\
        0 \quad\text{otherwise}
    \end{cases}.
\end{equation*}

\section{Problem Description}

\subsection{The DEC concept}

\begin{figure}[t]
	\centering
	\begin{tabular}{c}
		\IfFileExists{decloops.pdf}{\includegraphics[width=\columnwidth]{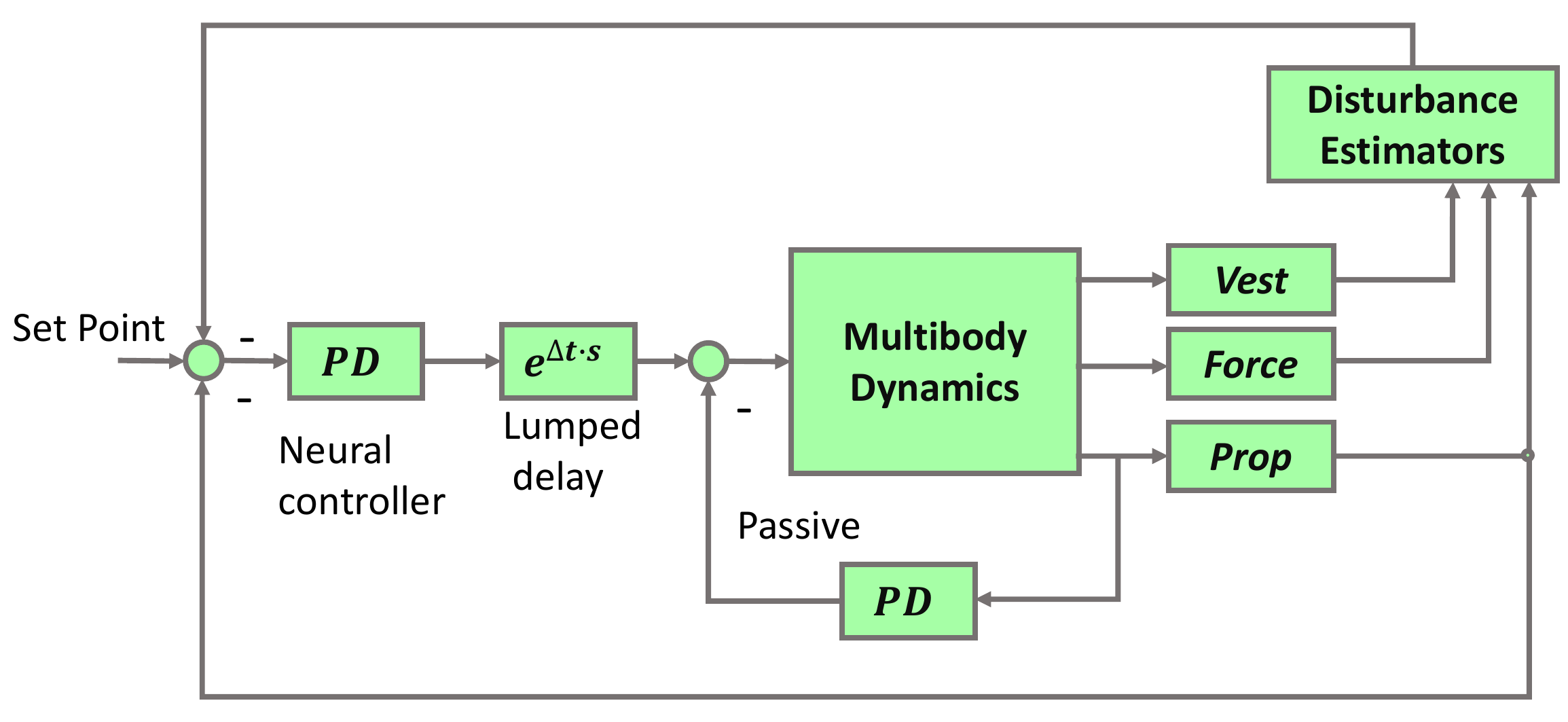}}{\includegraphics[width=\columnwidth]{Figures/decloops.pdf}}
		\vspace{10px}\\
		\IfFileExists{Fig1anglescrop.pdf}{\includegraphics[width=\columnwidth]{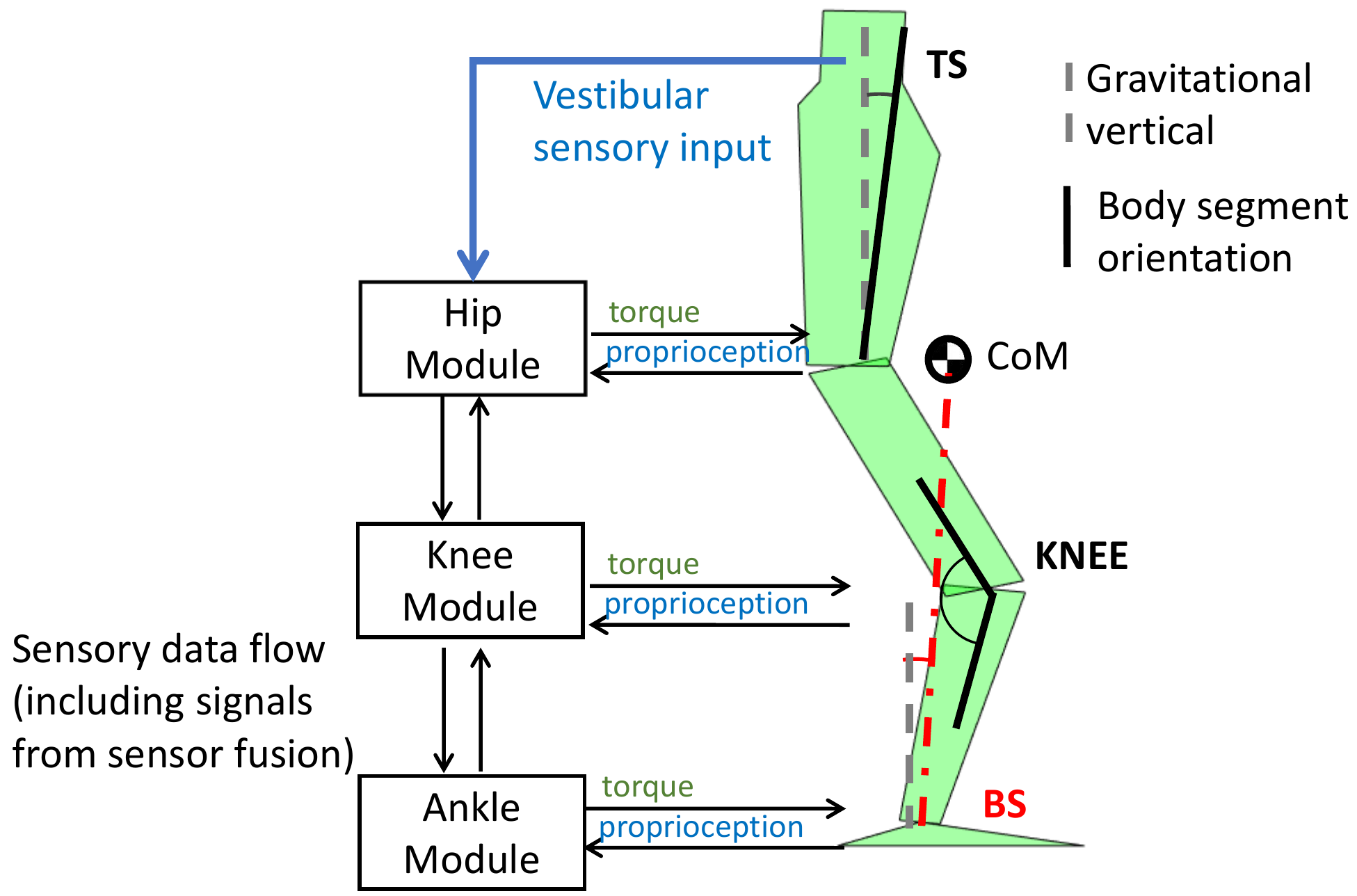}}{\includegraphics[width=\columnwidth]{Figures/Fig1anglescrop.pdf}}
	\end{tabular}			
	\caption{The DEC controller with a schematic model of the DEC concept (above) and the modular 3DoF control architecture used in the experiments (beneath). The angles used in the text are also displayed:  {TS} is the orientation of the trunk respect to the gravitational vertical; KNEE is the angle of the knee joint; BS is the angular sway of the CoM around the ankle joint and respect to the gravitational vertical. The \textit{lumped delay} accounts for all the delay effects that are distributed in general. }
	\label{DEC}
\end{figure}

The DEC concept provides a descriptive and predictive model of how human postural control mechanisms interact with movement execution control in producing a desired movement \cite{Mergner2010,10.1007/978-3-319-46669-9_42}. A schema of the DEC control is shown in Fig. \ref{DEC} and can be summarized as follows: 
\begin{itemize}
	\item A servo control loop for each degree of freedom (DoF). The servo is implemented as a PD controller, it is addressed in Fig. \ref{DEC} (above) as \textit{neural controller}. The controlled variable consists either of the joint angle, the orientation in space of the above joint, or the orientation in space of the center of mass of the whole body above the controlled joint.  {Variables are reconstructed locally using the exchanged sensory input};
	
	\item Multisensory estimation of the physical factors affecting the servo. These disturbances are \textit{rotation} and \textit{translation} of the supporting link or support, contact forces (e.g., push) and field forces (e.g., gravity impacting the supported link). Sensory channels are shown in Fig. \ref{DEC} as \textit{Vest},	
	\textit{Prop}, and \textit{Force}, representing, respectively, the vestibular (IMU), proprioceptive (encoders), and force (joint torque sensors) inputs;
	
	\item The disturbance estimates are fed into the servo so that the joint torque compensates on-line for the disturbances while executing the desired movements. 
\end{itemize}

The \textit{lumped delay} in Fig. \ref{DEC} (above) 
accounts for all the delay effects that, in general, are distributed. 
In particular, in robots, 
the main sources of delay 
are sampling rates 
in both sensors and computer-controlled system, see \cite{lippi2016human}. 
In humans, different control loops 
(e.g., proprioceptive feedback and disturbance compensation)
are associated with different delays (see \cite{antritter2014stability}) 
and the transport time within {the} 
nervous system creates differences 
in delay due to the {different distances that the neural 
signal has to cover in order to reach different joint positions}, 
i.e. lumped delay is estimated to be $180 ms$ for ankle joint control and $70 ms$ for hip joint control, see \cite{G.Hettich2014}. 
The disturbance compensation mechanism allows the system to maintain a low loop gain and thus stable control in face of neural time delays. 
A further, indirect limitation to the gain is represented 
by the maximum torque that the foot 
can produce on the ground without losing contact. 
The DEC concept can be generalized to a modular control architecture, 
where estimators in each module treat 
disturbances acting on all supported links as 
if affecting a single inverted pendulum, see \cite{V.Lippi2013}. 
This approach has been applied to multiple DoF robots \cite{10.3389/fnbot.2017.00049,lippi2016human,
	ZebenayLippiMergener2015,lippi2018prediction}. 
The reference input to each module 
determines its postural function, 
e.g. maintaining a given orientation 
of the supported link (either in space or with 
respect to the supporting link), 
or maintaining the center of mass (CoM) 
above its supporting joint. 
Modules exchange information 
with neighboring 
modules, i.e. those mechanically interconnected.

\subsection{Modularity, Coupling Forces and Delays}
Since, in DEC control, 
each DoF of the humanoid is controlled by one DEC module,
this results in \textit{control modularity}. 
Each module commands the torque to be applied to the controlled DoF. 
In previous work on the topic, 
the desired trajectory is specified as an input to each module, 
specifically in the form of a 
reference for a desired variable, i.e. a joint angle, an orientation in space of the link supported by the controlled joint, or the orientation in space of the center of mass of the whole set of links above the supported joint (e.g. for the ankle joint it would be BS in Fig. \ref{DEC}), see
\cite{lippi2016human}. 

The multitasking capability of the DEC control 
consists of using each DoF to perform a 
different task, see \cite{Lippi2015}. 
All modules have the same structure 
and there is no centralized model of the whole system. 
Modules operate not completely independently of each other, 
since they exchange sensory information, 
i.e. coordinate transformation across the joints 
that interconnect the body segments. 
The DEC model can be defined as a \textit{low level 
control system} taking
care of the fundamental task 
of posture control and acting at the level of joint kinematics. 
Coordination between joints emerges from the 
interaction between different modules 
and between the modules and the body mechanics. 
At the level of modules, no kinematic synergy is explicitly specified.

\begin{figure}[htbp]
	\centering
	\IfFileExists{swingplotNoBid1.pdf}{\includegraphics[width=\columnwidth]{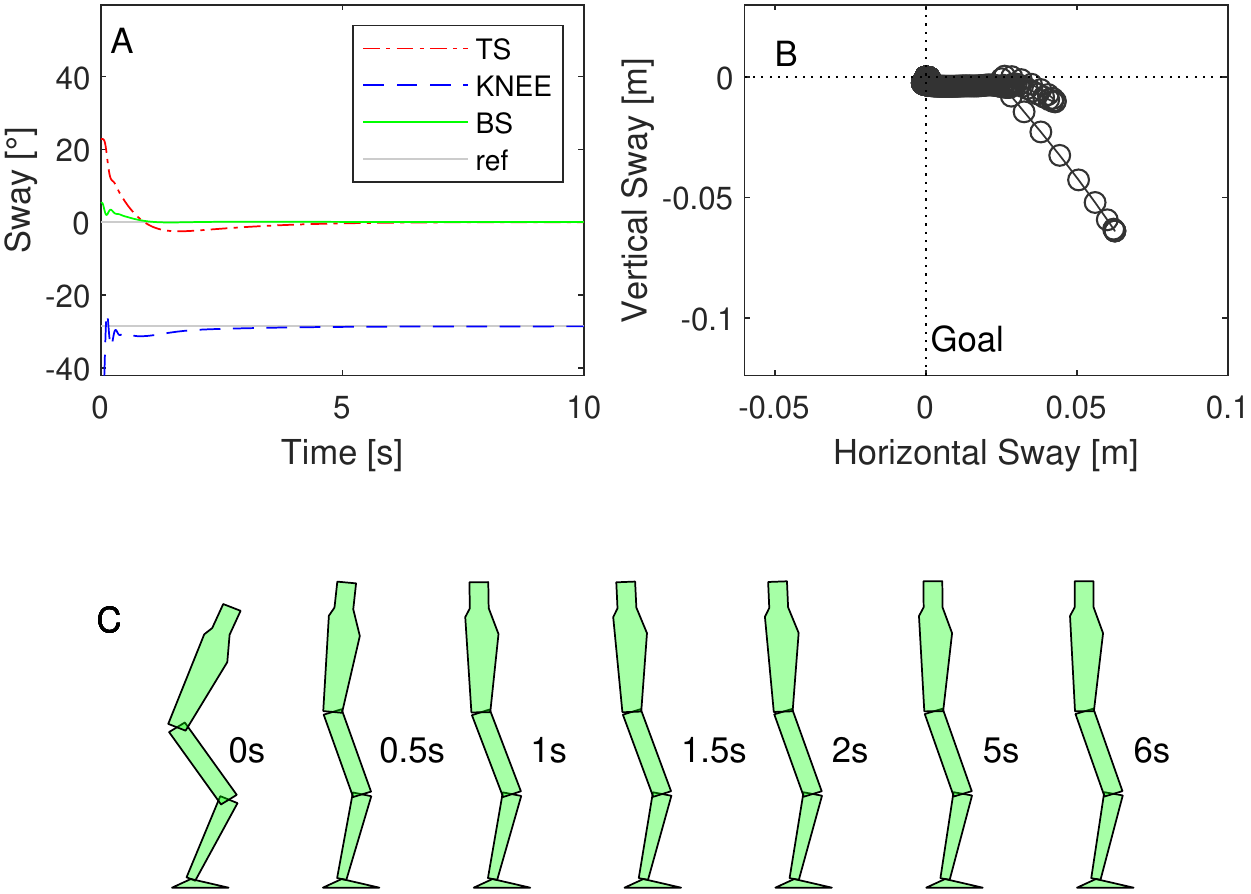}}{\includegraphics[width=\columnwidth]{Figures/swingplotNoBid1.pdf}}
	\caption{Transient behavior of body posture in response to a movement commanded from a displaced position. In (A) the evolution of the orientation of the three body segments in space is shown (TS = trunk in space, THS = Thigh in space, SS = Shank in space). In (B) the trajectory of the center of mass in the sagittal plane is shown (each dot represents a sample taken at 100 Hz). Notice the typical circular movement, qualitatively resembling the behavior described in \cite{ott2016good} for the modular control system, with modules that can produce conflicting commands. In (C) the body trajectory is shown as a succession of body poses.}
	\label{fig:Overshoot}
\end{figure}

In \cite{ott2016good}, it is shown how 
the modular structure of the DEC can 
lead to conflicts between modules: 
an example is the circular 
overshoot exhibited in body posture's 
transient behavior, as shown in Fig. \ref{fig:Overshoot}. 
Specifically, the knee module 
commands an extension of the leg. 
As the upper body is also perturbed forward, 
an extension of the knees produces a disturbance 
for the ankle joints, 
which try to move the CoM back 
to the vertical equilibrium position. 
In general, coupling forces between 
body segments are a challenge for distributed modular controllers,
especially considering the presence of delays, see e.g. \cite{LippiMergner2015}. 
In Fig. \ref{fig:Overshoot}, 
the transient behavior 
of body posture is shown. 
Similarly to \cite{ott2016good}, 
the control parameters and the joint passive 
stiffness were chosen to produce 
a compliant behavior to 
emphasize the effect of competing controllers.  
Nevertheless, the used parameters are realistic 
in the sense that are able to stabilize the body. 
The absence of passive stiffness is common 
in humanoids robots actuated with DC motors. 
The parameters used in the simulation are 
reported in Table \ref{paramTAB}. 

It should be noticed that the original formulation of the DEC model was designed to describe steady state behavior in human subjects. The distributed control proposed in this work is thought as a possible solution to humanoid control. The comparison with human experiments is beyond the scope of this work.  

\begin{table}[htb]
\vspace{10px}
\centering
\label{paramTAB}
\begin{tabular}{|l|l|rl|}
\hline
Segment/joint & Parameter & value &  \\ \hline
Trunk/hip & $K_p$ & 73.57 & N*m/rad \\ \cline{2-4} 
 & $K_d$ & 18.394 & N*m*s/rad \\ \cline{2-4} 
 & mass & 30 & Kg \\ \cline{2-4} 
 & length & 0.5 & m \\ \cline{2-4} 
 & center of mass & 0.25 & m \\ \cline{2-4} 
 & kp passive & 0 & N*m/rad \\ \cline{2-4} 
 & kd passive & 0 & N*m*s/rad \\ \cline{2-4}
 & $G_{servo}$ & 1 &   \\ \cline{2-4}
 & Lumped Delay & 10 & ms \\\hline
Thigh/knee & $K_p$ & 220.72 & N*m/rad \\ \cline{2-4} 
 & $K_d$ & 16.55 & N*m*s/rad \\ \cline{2-4} 
 & mass & 10 & Kg \\ \cline{2-4} 
 & length & 0.5 & m \\ \cline{2-4} 
 & center of mass & 0.25 & m \\ \cline{2-4} 
 & $k_p$ passive & 0 & N*m/rad \\ \cline{2-4} 
 & $k_d$ passive & 0 & N*m*s/rad \\ \cline{2-4}
 & $G_{servo}$ & 1 & N*m*s/rad \\ \cline{2-4}
 & Lumped Delay & 10 & ms \\\hline
Shank/ankle & $K_p$ & 465.98 & N*m/rad \\ \cline{2-4} 
 & $K_d$ & 116.49 & N*m*s/rad \\ \cline{2-4} 
 & mass & 10 & Kg \\ \cline{2-4} 
 & length & 0.5 & m \\ \cline{2-4} 
 & center of mass & 0.25 & m \\ \cline{2-4} 
 & $k_p$ passive & 0 & N*m/rad \\ \cline{2-4} 
 & $k_d$ passive & 0 & N*m*s/rad \\ \cline{2-4}
 & $G_{servo}$ & 1 &   \\ \cline{2-4}
 & Lumped Delay & 10 & ms \\\hline
\end{tabular}
\caption{Control and simulation parameters}
\end{table}

\subsection{Distributed control problem}
The main contribution of this work
is the design of a distributed control approach,
which harnesses the modular nature of the DEC control
while trying to reduce conflicts among joints. This approach is innovative in the field of bio-inspired humanoid posture control.
The underlying idea is to enable 
only one module at a time
to change its positional reference.
This strategy claims to prevent
the circular overshoot of  {the} CoM trajectory of Fig. \ref{fig:Overshoot}.
Starting at time $t=0$,
every $T_e\in\mathbb{R}_{>0}$ seconds,
a new control module is enabled
and all the others are disabled.
Precisely, disabled modules implement only 
the gravity compensation control on torque and
are controlled to be at the current fixed position, as it will
be explained in Section \ref{sec:posContrWGrav}.
Switching between controlled modules
would traditionally require a central decision.
However,
the absence of any centralized control structure
results in the need of an agreement among modules.
Traditionally, in those cases where
a multi-agent system 
is seeking an agreement
on a variable of common interest,
consensus-based strategies are employed, see 
\cite{ren2005survey}.
In Section \ref{sec:maxCons},
the mentioned consensus protocol is analyzed.
Consensus has been used in
robotics for inter-robot and inter-vehicular coordination, 
see \cite{notarstefano2006distributed} (rendezvous),
\cite{brunet2008consensus} (task assignment),
or \cite{automation2018Molinari} (conflict resolution); in this paper,
the paradigm changes, since consensus is here exploited
for intra-robot coordination.
This approach is suitable for
plug-and-play  modules
where initialization is not required, 
since the parameters of each module are defined independently of the ones of the other modules, only on the basis of body anthropometrics. 

\section{Control Design}
\subsection{Posture Control Scenario}
Humanoid balance in the sagittal plane can be modeled as the control of a multiple inverted pendulum by means of joint torques, on the basis of sensor inputs, i.e. encoders and  {inertial measurement units}. 
The body is modeled as a triple inverted pendulum, following the robot configuration used in the robotic experiment presented in \cite{ott2016good} where ankle, knee and hip were actuated. The model used in the simulation is implemented in Matlab/Simulink, and it is the same used in \cite{alexandrov2005feedback} and \cite{V.Lippi2013}.  

\subsection{Control implementation}
Let $\mathcal{M}$ be the set of all modules.
They can exchange information
via a wired communication network.
Under a graph-theoretical point of view, 
let the directed graph $(\mathcal{M},\mathcal{A})$
model the wired network topology,
with $\mathcal{M}$ being the set of nodes and
$\mathcal{A}\subset\mathcal{M}\times\mathcal{M}$ the respective set of arcs.
For any pair $i,j\in\mathcal{M}$,
$(i,j)\in\mathcal{A}$
if node $i$ receives information from node $j$.
\begin{assumption}
	\label{ass:retrieveInf}
	Each module can retrieve information
	from all the modules connected to it
	by a body segment (see Figure~\ref{DEC}).
\end{assumption}
By Assumption \ref{ass:retrieveInf}, 
it can be easily shown that
$(\mathcal{M},\mathcal{A})$ presents a 
connected topology (further details in \cite[A Tutorial on Graph Theory]{ren2007information}).
In the following,
$\forall i\in\mathcal{M}$,
let $N_i\subseteq\mathcal{M}$ be 
the set of modules sending information
to node $i$, i.e.
\begin{equation}
	N_i=
	\{
		j\in\mathcal{M} \mid
		(i,j)\in\mathcal{A}
	\}.
\end{equation}

We consider {the problem of} controlling body posture and equilibrium in the body sagittal plane. The body is represented as a triple inverted pendulum standing on a fixed support surface. The state of the system is described by the joint angle of ankles, hips and knees, or, equivalently, by the orientation in space of body segments or CoM. In particular, each module, say $i\in\mathcal{M}$, is associated to a different task, meaning that it is controlling a specific variable $\alpha_i$:
\begin{itemize}
\item  the \textit{ankle module} controls the body in space variable, $\alpha_i= BS$, i.e. the CoM sway respect to the foot;
\item the \textit{knee module} controls the knee joint angle $\alpha_i= KNEE$;
\item the trunk is controlling the trunk orientation in space $\alpha_i=TS$.
\end{itemize}
In Figure~\ref{DEC}, the general 
structure of each controller 
is shown above and the relationship 
between modules is shown beneath. 
The controlled variable $BS$ 
is constructed in the \textit{ankle module} 
using the down-channeled signal.

\begin{figure*}[htbp]
	\centering
	\IfFileExists{SwithchC.pdf}{\includegraphics[width=0.8\textwidth]{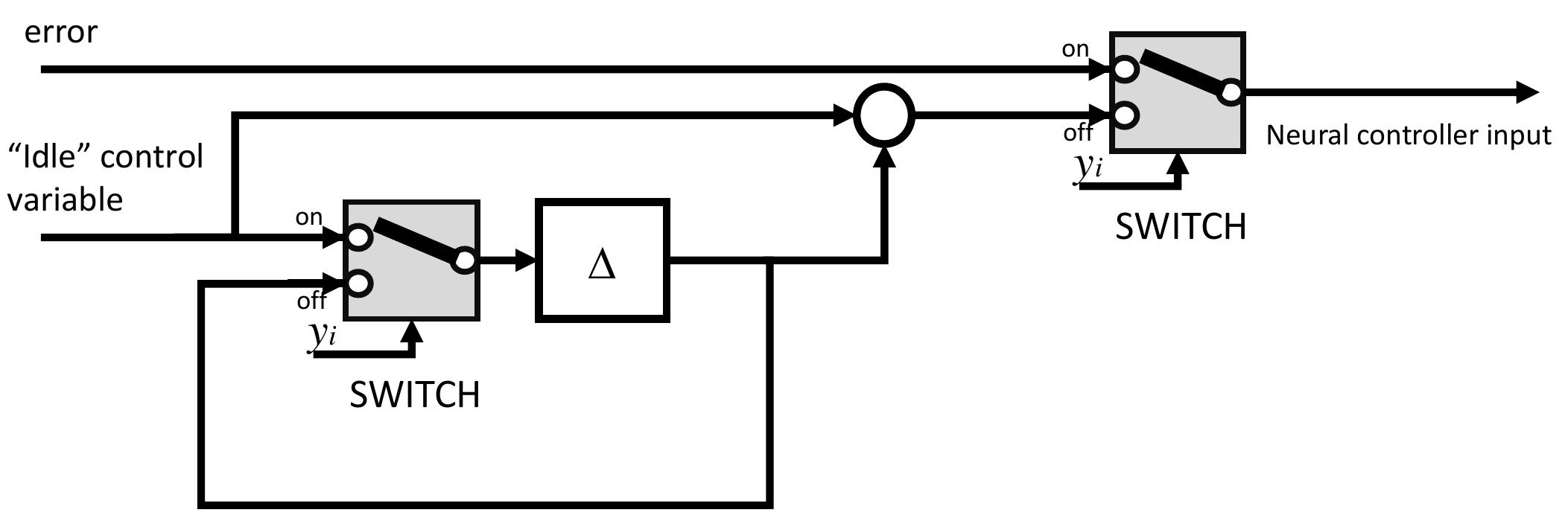}}{\includegraphics[width=0.8\textwidth]{Figures/SwithchC.pdf}}
	\caption{Switching between \textit{enabled} and \textit{disabled} mode for a control module. When the module is set on \textit{enabled} the error on the controlled variable is fed as input to the neural controller (PID). A second control variable, i.e. is kept in memory by a register/delay (block $\Delta$). When the state $y_i$ of the module switches to \textit{disabled} the PID controller is commanded to keep constant such variable. The value used as reference for the \textit{disabled} mode is a variable describing the state of the controlled link before swithcing from \textit{enabled} mode. In the presented example the orientation is space of the links will be used.}
	\label{fig:SwithchC}
\end{figure*}

Let, 
\begin{equation}
	\label{eq:authVar}
	\forall i\in\mathcal{M},\
	\forall k\in\mathbb{N}_{\geq0},\
	y_i(k)\in\{0,1\}
\end{equation} 
be a binary variable,
referred to as \textit{enabling variable},
defined as follows:
module $i$ is enabled
in the continuous time interval $$I_k:=(kT_e,\ (k+1)T_e],$$
$k\in\mathbb{N}_{\geq0}$
and $T_e\in\mathbb{R}_{>0}$ a real-valued time,
if and only if $y_i(k)=1$.
Moreover,
\begin{equation}
	\label{eq:exclusiveY}
	\forall k\in\mathbb{N},\ 
	\exists!i\in\mathcal{M}:\ y_i(k)=1.
\end{equation}
The enabling variables are initialized as,
$\forall i\in\mathcal{M},\
y_i(0)=1$. 
{Disabled} 
modules are controlled to the current position,
see Figure~\ref{fig:SwithchC}.

Modules run
a consensus protocol
which let them agree
on which one is the enabled module.
At each time $t=kT_e$, $k\in\mathbb{N}$,
each module has a state $w_{i_0}^k\in\mathbb{R}_{\geq0}$,
that quantifies the 
need for
that module 
to be enabled.
The value of $w_{i_0}^k$ is defined as the error on the controlled variable, i.e.
\begin{itemize}
	\item the CoM sway for the ankle joint; 
	\item the knee joint angle for the knee joint;
	\item the trunk sway for the hip joint.
\end{itemize}
Intuitively, the enabled module 
during interval $I_k$
will be the one retaining the highest 
$w_{i_0}^k$.
That is to say, $\forall k\in\mathbb{N}$,
\begin{equation}
	\label{eq:chooseY}
	y_i(k)=1 
	\iff
	i = 
	\argmax_{j\in\mathcal{M}}
	\left(
		w_{j_0}^k
	\right).
\end{equation}
\begin{assumption}
	\label{ass:diffBet}
	$\forall i,j\in\mathcal{M},\
	i\not=j$, $w_{i_0}^k\not=w_{j_0}^k$.
\end{assumption}
Clearly, by Assumption \ref{ass:diffBet},
(\ref{eq:chooseY})
implies (\ref{eq:exclusiveY}).
One way 
nodes can achieve the solution
in (\ref{eq:chooseY}) 
-- without the presence of
any central element -- is by 
running a distributed max-consensus protocol.

\subsection{Max Consensus}
\label{sec:maxCons}
In the following,
a discrete-time max-consensus protocol is presented,
which will iterate at every instant $kT_e$, $k\in\mathbb{N}$,
thus allowing for distributively retrieving $y_i(k)$, $\forall i\in\mathcal{M}$.
Initially,
all modules have their respective $w_{i_0}^k$.
Let an iteration variable be defined for
each module $i\in\mathcal{M}$ 
as $w_i^{(k)}:\mathbb{N}_{\geq0}\mapsto\mathbb{R}_{\geq0}$,
such that $w_i^{(k)}(0)=w_{i_0}^k$.
All modules iterate the following protocol:
\begin{equation}
	\label{eq:consProt}
	\forall\kappa\in\mathbb{N}_{\geq0},\ 
	w_i^{(k)}(\kappa+1)=\max_{j\in N_i\cup\{i\}} w_j^{(k)}(\kappa),
\end{equation}
where $\kappa$ denotes the iteration index.
\begin{proposition}
	Given a connected network topology
	$(\mathcal{M},\mathcal{A})$,
	if all modules in $\mathcal{M}$
	iterate (\ref{eq:consProt}),
	then consensus is achieved at $\bar{\kappa}\in\mathbb{N}$, such that
	\begin{equation}
		\label{eq:achievedMaxCons}
		\forall\kappa>\bar{\kappa},\ 
		\forall i\in\mathcal{M},\
		w_i^{(k)}({\kappa})=\max_{j\in\mathcal{M}}w_j^{(k)}(0):=w^*.
	\end{equation}
	\begin{proof}
		Protocol (\ref{eq:consProt})
		is a traditional max-consensus protocol.
		By \cite{nejad2009max},
		consensus in connected 
		network topologies is reached
		on the max-value in a number of steps
		depending only on the network topology.		
		By Assumption \ref{ass:retrieveInf}, 
		the communication network topology is connected,
		therefore max-consensus is achieved				
		in the sense of (\ref{eq:achievedMaxCons}).
		Moreover, by \cite{nejad2009max},
		with the given topology,
		$\bar{\kappa}=2$.
	\end{proof}
\end{proposition}
As soon as consensus is achieved,
modules can compute their respective $y_i(k)$ 
as follows:
\begin{equation}
	\label{eq:computeY}
	\forall i\in\mathcal{M},\
	\forall k\in\mathbb{N},\ 
	y_i(k)=\mathbb{I}\left(
		w^k_{i_0}=w^*
	\right).
\end{equation}
\begin{proposition}
	Under Assumption \ref{ass:diffBet},
	(\ref{eq:computeY}) implies (\ref{eq:exclusiveY}).
	\begin{proof}
		By Assumption \ref{ass:diffBet},
		there is only one module, say $i^*\in\mathcal{M}$,
		such that
		$w_{i^*_0}^k=w^*$.
		By (\ref{eq:computeY}),
		$$ 
		\forall j\in\mathcal{M}\setminus\{i^*\},\ 
		y_j(k)=\mathbb{I}\left(
		w^k_{j_0}=w_{i^*_0}^k
		\right) 
		=0,
		$$
		from which (\ref{eq:exclusiveY}) immediately follows.
	\end{proof}
\end{proposition}

As mentioned above, this
solution is compatible with a plug-and-play framework.
In this context, letting modules
communicate over a wireless network
can speed up the set-up 
of the system.
However, traditionally,
wired communication is sensibly 
faster than the wireless one.
Convergence speed of the consensus protocol (\ref{eq:consProt})
can be improved in the wireless framework
by using the strategy presented in 
\cite{MOLINARI2018176}.

\section{Validation Experiment}
\label{sec:posContrWGrav}
The system is tested with the task shown in Fig. \ref{fig:Overshoot}. The parameters are shown in Table \ref{paramTAB}.
They are the same for both the presented cases. 
In particular, passive stiffness and damping 
have been set to $0$ and the delay to $10 ms$, 
i.e. a small delay (compared, for example to the 
$180 ms$ presented in \cite{G.Hettich2014}) 
that anyway poses realistic limitations on 
the servo controller gain. This choice was 
made because in this work we do not want to 
examine the relationship between delay and 
passive stiffness studied in \cite{ott2016good},
but we want to emphasize the relationship between 
competing modules. 
In Figure~\ref{DEC}, both the 
disturbances estimators and the 
proprioceptive signals are fed as 
inputs to the neural controller. 
This is performed by setting the proportional gain $K_p$ to $mgh$, where $m$ is the mass of the body above the controlled joint, $g=9.81 m/s^2$ is the gravity acceleration and $h$ the height of the CoM. The derivative component is set to a fraction of the proportional one. This way, the gravity error is expressed as the CoM sway angle and the other estimators are expressed as an ``angle equivalent'', in the sense that the desired corrective torque is divided by $mgh$. 
This implies that, for the controller and all the compensated disturbances,
the ratio between $K_p$ and $K_d$ is fixed, 
while each signal can be associated with a specific gain. 

In this work, the DEC has been implemented as shown in \cite{ott2016good} with a separate neural controller for each signal. A PID controller is designed for the servo and a PD controller for disturbance compensation (with gravity compensation an integrative action is not desired). Body segment positions and velocities are assumed to be known exactly. Since we consider only gravity as external disturbance, the control torque is expressed by:

\begin{equation}
	\forall i \in \mathcal{M}, \ \tau_i=G_g\alpha_{CoM_i}+G_{servo}\epsilon_i (K_p+ sK_d) e^{s \Delta t} 
\end{equation}
where $\epsilon_i$ is the error on the controlled variable and $\alpha_{CoM_i}$ the angle of the CoM with respect to the controlled joint.
The used $\epsilon_i$ depends on $y_i$ as follows:
\begin{equation}
	\epsilon_i=\left\{ \begin{array}{lr}
		\alpha_i-\alpha^{ref}_i & \mathrm{if }\ y_i=1 \\
		\alpha_i-\bar{\alpha}^{ref}_i & \mathrm{if }\ y_i=0
\end{array}
\right. ,
\end{equation}
where $\alpha_i$ is the respective controlled variable. The value $\bar{\alpha}^{ref}_i$ is set to the value of $\alpha_i$ at the instant of deactivation, when $y_i$ makes a transition from $1$ to $0$, as modeled by the block $\Delta$ in Fig. \ref{fig:SwithchC}.

\section{Results}
\begin{figure*}[htbp]
	\centering	
	\IfFileExists{swingplot1.pdf}{\includegraphics[width=0.6\textwidth]{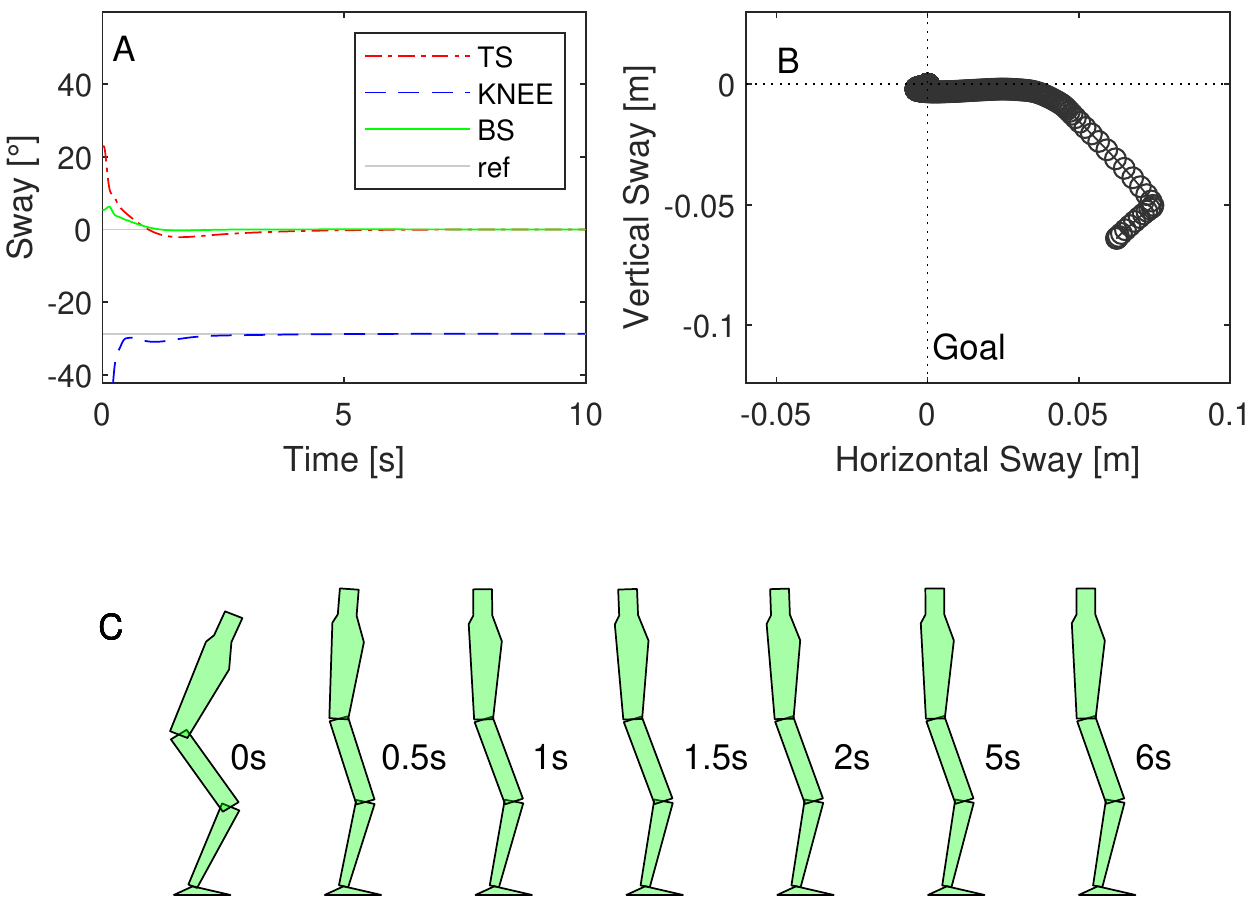}}{\includegraphics[width=0.6\textwidth]{Figures/swingplot1.pdf}}
	\caption{Transient behavior of body posture using the distributed control strategy. The starting position and the reference are the same as the example shown in Fig. \ref{fig:Overshoot}. The trajectories shown in (A) are comparable to the ones shown in Fig. \ref{fig:Overshoot}.In (B)  {it} is possible to notice that the system is not producing a circular movement of the CoM but rather straight lines produced by the activities of the single modules. In (C) the body trajectory is shown as a succession of body poses.}
	\label{fig:Results}
\end{figure*}

\begin{table}[t]
	\vspace{10px}
	\centering
	\begin{center}
		\begin{tabular}{|l|l|l|l|}
			\hline\rule{0pt}{3ex}    
			Variable & Index & Original & Distributed \\
			\hline
			\hline\rule{0pt}{3ex}    
			TS & overshoot & 2.5118\textdegree &  2.1166\textdegree \\
			& rise time & 0.80 s & v 0.84 s\\
			& settling time & 9.99 s & 9.99 s \\
			\hline\rule{0pt}{3ex}    
			KNEE & overshoot & 3.4765\textdegree & 0\textdegree\\
			& rise time & 0.07 s & 0.31 \\
			& settling time & 9.99 s &  9.99 s \\
			\hline\rule{0pt}{3ex}    
			BS & overshoot & 0.0961\textdegree & 0.3075\textdegree\\
			& rise time & 0.81 s & 0.86 s\\
			& settling time & 9.99 s &   9.99 s\\ 	
			\hline\rule{0pt}{3ex}    
			& energy & 196.72 J& 68.25 J \\
			\hline
		\end{tabular}
	\end{center}
		\caption{Dynamic performance}
		\label{performance}
\end{table}
In order to evaluate the impact of 
the designed distributed control strategy,
the transient behavior of the DEC Control is compared, 
with and without distributed control policy, in response to a sudden change of reference. 
The results are shown, respectively, in Figure~\ref{fig:Overshoot}  
and Figure~\ref{fig:Results}. 
Performances are summarized in Table \ref{performance}.

The dynamic performances for the two controllers are comparable. 
The CoM trajectory does not produce any circular 
(or overshooting)  movement
but it is rather described by straight lines,
clearly
due to the switching behavior.
While for TS and KNEE there is
a substantial drop of the overshoot measure, 
a slight increase of this measure affects BS. 
As it clearly emerges from the table, 
the cost to pay for a decreased overshoot 
is an increase of the rise time. 
Energy is intended as the integral 
of the mechanical power provided 
at the joints; on a real robot,
the power consumption can be heavily 
influenced by the actuation 
(e.g. DC motors require power in order to hold static positions). 
In this scenario, the max-consensus 
algorithm makes the DEC control more energy efficient. 
\section{Discussion}
This work has discussed 
a distributed control approach for 
the modular bio-inspired DEC controller,
where modules negotiate their authorization to move.
The DEC original formulation (in Figure~\ref{fig:Overshoot}) 
is compared to the designed distributed control strategy (in Figure~\ref{fig:Results}).

The simulated humanoid was initialized from an initial
position, from which it had to reach the upright pose;
trajectories of body segments and 
of the CoM position in the sagittal plane were recorded.  
Traditionally,
DEC system exhibits a mutual obstruction between modules, 
resulting in a circular (overshooting) CoM trajectory. 

This work's novelty lies in the fact that modules 
distributively 
agree on a common strategy.
By doing so,
within a DEC framework,
conflicts between modules are avoided by 
having only one of the modules \textit{enabled} at a time.
With the addition of this distributed agreement strategy, 
the transient response does not show circular CoM trajectories anymore (see Figure~\ref{fig:Results}). 
Moreover, our proposed distributed control strategy 
appears to be more energy efficient. 
Such improvements come at expenses of a small 
delay on the settling time, due to the modules' inactivity when disabled.




\section*{ACKNOWLEDGMENT}
We gratefully acknowledge financial support for the project MTI-engAge (16SV7109) by BMBF.

This work was also funded by the German Research Foundation (DFG) within
their priority programme SPP 1914 ”Cyber-Physical Networking (CPN)”,
RA516/12-1.


\bibliographystyle{IEEEtran}
{\small
\bibliography{vittorio}}

\end{document}